\documentclass[aps,prb,twocolumn,superscriptaddress]{revtex4-1}

\usepackage{graphicx}
\usepackage{xcolor}
\usepackage{dcolumn}
\usepackage{bm}
\usepackage{amsmath}
\usepackage{amssymb}

\usepackage[utf8]{inputenc}
\usepackage[T1]{fontenc}
\usepackage{mathptmx}
\usepackage{etoolbox}

\makeatletter
\def\@email#1#2{%
 \endgroup
 \patchcmd{\titleblock@produce}
  {\frontmatter@RRAPformat}
  {\frontmatter@RRAPformat{\produce@RRAP{*#1\href{mailto:#2}{#2}}}\frontmatter@RRAPformat}
  {}{}
}%
\makeatother

\begin{document}

\title{Spiral, target, stripe, and disordered waves in active six-state Potts models}

\author{Hiroshi Noguchi}
\email{noguchi@issp.u-tokyo.ac.jp}
\affiliation{Institute for Solid State Physics, University of Tokyo, Kashiwa, Chiba 277-8581, Japan}
\date{\today}

\begin{abstract}
Wave propagation can be observed in various nonequilibrium systems.
In this study, we investigated the properties of several wave modes in active six-state Potts models using Monte Carlo simulations of square and hexagonal lattices. 
Disordered and spiral (SP) waves of six states are formed under weak and strong repulsions at nonflip contacts, respectively. 
The target (TG) and stripe (ST) waves were found to emerge under stronger repulsion. 
These three wave modes (SP, TG, and ST) can temporally coexist in small systems near the transition points but they do not switch in large systems or far from these transition points. During coarsening from randomly mixed states to ST waves, SP waves appear at an intermediate stage. 
The SP wave modes of three even- or odd-numbered states (states $s=0,2,4$ or $s=1,3,5$) emerge under two conditions:
repulsion at the diagonal contact and attraction at nonflip contacts.
Previously thought to be identical for both conditions, 
the wave types were found to differ, comprising forward and backward waves ($s=1\to 3\to 5\to 1$ or $s=1\to 5\to 3\to 1$), whose domain boundaries move by the two-step and four-step forward flips, respectively.  The transition between the waves of the even- and odd-numbered states is first-order for both the forward and backward waves. 
\end{abstract}

\maketitle

\section{Introduction}

Various spatiotemporal patterns appear away from thermal equilibrium.\cite{nico77,hake04,mikh94,kura84,murr03}
In particular, wave propagation is frequently observed.
These waves play a role in the transduction of biological signals, including in neuronal electrical signals, muscle contraction, cell migration, and differentiation.\cite{beta17,bail22,cher08,devr17,dene18}
In intracellular and other microscopic waves,
 thermal fluctuations can be significant;\cite{beta17,bail22,nogu24c} however, the effects of fluctuations and nucleations are not understood so far.

Lattice models have been widely used to study spatiotemporal patterns.
The lattice Lotka--Volterra models\cite{szol14,reic07,szcz13,kels15,dobr18,szab08,baze19,mene22,yang23,szol23,clif25} are based on self-reproduction and predator–prey interactions.
These models include noise via a random site selection; however, this noise is considered environmental rather than thermal.
Since the population increases through self-reproduction, the dynamic patterns disappear in the long-term limit owing to extinction.
Recently,  active Potts models\cite{nogu24a,nogu24b,nogu25,nogu25a,nogu25b,nogu26a,nogu26b} have been developed, which consistently incorporate thermal noise to locally satisfy the detailed-balance condition. Moreover, these models take nucleation into account and can sustain spatiotemporal patterns in the long-term limit.

The aim of this study is to clarify the wave properties in six-state active Potts models,
in which various types of wave modes emerge.\cite{nogu25b}
In particular, spiral (SP) waves, comprising three types of domains (even- or odd-numbered states),
propagate through nucleation and growth via the other three states.
The scaling exponents of the second-order transition between this three-state wave mode and the mixing modes 
differ from the equilibrium values.
In this study, we examined a wider range of parameters than in our previous studies\cite{nogu25b}
and investigated the temporal correlations and spatial structures.
Moreover, we clarified the boundaries of the wave modes and identified target (TG) and stripe (ST) waves as new modes [the abbreviation of dynamic modes are listed in Table.~\ref{tab:1}].

Previously, we studied the coarsening dynamics from a randomly mixed state to SP and nonspiral disordered (DS) wave modes.\cite{nogu26b}
Here, we investigated the coarsening dynamics to the ST wave mode.
In diffusion-dominated conserved systems,
a characteristic length $\ell$ in coarsening dynamics toward an equilibrium state increases in accordance with the  Lifshitz--Slyozov law ($\ell \propto t^{1/3}$).\cite{lifs61}
However, the hydrodynamic and elastic interactions can change this scaling exponent.\cite{puri09,furu85,bray94,tana00}
In particular, Hajime Tanaka and his coworkers revealed that elastic network formation during the coarsening leads to
 $\ell \propto t^{1/2}$.\cite{tana00,tana97,tate21,yuan25}
Recently, coarsening dynamics leading to nonequilibrium states have received growing attention from active-matter researchers.
In the diffusion-dominated conserved systems,
self-propelling activity has been shown to induce coarsening exponents that are either higher\cite{mish14} or lower\cite{saha20,katy20,patt21} 
than $1/3$.
In nonconserved systems, such as Ising and Potts models,
surface-tension-driven coarsening into an equilibrium state exhibits $\ell \propto t^{1/2}$, following the Lifshitz--Allen--Cahn (LAC) law.\cite{lifs62,alle79} 
However, coarsening dynamics leading to nonstatic states have been far less explored in nonconserved active systems.

The models and methods are described in Sec.~\ref{sec:model}.
Simulation results are presented and discussed in  Sec.~\ref{sec:results}.
The dynamic modes and phase diagrams are briefly explained in Sec.~\ref{sec:rev}.
The wave modes comprising the domains of all six states are examined in  Sec.~\ref{sec:w6}.
The coarsening dynamics originating from a random mixture are presented in Sec.~\ref{sec:cd}.
The SP wave modes consisting of three domain types are examined in  Sec.~\ref{sec:w3}.
The effects of lattice type are discussed  in  Sec.~\ref{sec:hex}.
Finally, a summary is presented in Sec.~\ref{sec:sum}.

\begin{table}
\caption{\label{tab:1} 
Abbreviation of dynamic modes}
\begin{center}
\begin{tabular}{ll}
 \hline
W$n$ & wave comprising $n$ types of domains \\
WI   & intermediate wave mode\\
M$n$ & $n$-state mixing phase \\
HC$n$ & homogeneous cycling of $n$ phases \\ 
SP wave & spiral wave \\
TG wave & target wave \\
ST wave & stripe wave \\
DS wave & disordered wave \\
W3f & three-state forward wave \\
W3b & three-state backward wave  \\
M2HC3 & homogeneous cycling of two-state mixing phases \\
M2W3 & wave comprising two-state mixing phases \\
\hline
\end{tabular}
\end{center}
\end{table}

\section{Models and Methods}\label{sec:model}

In a $q$-state Potts model, each site possesses one of the $q$ states ($s\in [0,q-1]$).
The nearest neighboring sites ($i$ and $i'$) have contact energies $J_{s_i,s_{i'}}$ as in equilibrium Potts models.\cite{wu82,pott52}
The interaction energy $H_{\mathrm{int}}$ is given by
\begin{equation}
\label{eq:hint}
H_{\mathrm{int}} = - \sum_{\langle ii'\rangle} J_{s_i,s_{i'}}.
\end{equation}
In equilibrium systems, each state may additionally possess self-energy $\varepsilon_s$,
and the ratio of the forward and backward flip rates for the flipping of a single site from $s$ to $s'$ is given by $\exp(-\Delta H_{s_is'_i})$, where $\Delta H_{s_is'_i} = \Delta H_{\mathrm{int}} - h_{s,s'}$ is the energy difference between the two states and 
$h_{s,s'} = \varepsilon_s - \varepsilon_s'$ is the flip energy.
In this paper, the thermal energy $k_{\mathrm{B}}T$ is normalized to unity.
At equilibrium, the sum of flip energies along a flip cycle vanishes; 
thus, $\sum_k h_{k,[k+1]}=0$, where $[k']$ is $k' \bmod q$.
We extended the Potts models to a nonequilibrium framework with nonzero cyclic sums while maintaining $h_{ss'}=-h_{s's}$.\cite{nogu24a}
Therefore, the detailed balance cannot be globally satisfied, although the local balance condition holds for flips between neighboring states.
For three-state cycles, this corresponds to the rock--paper--scissors relationship.
These types of dynamics can be realized in reactions on a catalytic surface\cite{ertl08,bar94,goro94,barr20,zein22} and molecular transport across membranes.\cite{tabe03,miel20,holl21,naka18,nogu23}
Three intermediate products and five membrane binding states exist in six-state cycles, respectively.
Active Potts models with global coupling were studied by Esposito et al.\cite{herp18,meib24,ptas25} They called them driven Potts\cite{herp18,meib24} and nonequilibrium Potts models.\cite{ptas25}
The four-state active vector-Potts model is also called the nonreciprocal Ising\cite{avni25} and nonreciprocal Ashkin--Teller model.\cite{nogu25b}

\begin{figure}[tbh]
\includegraphics[]{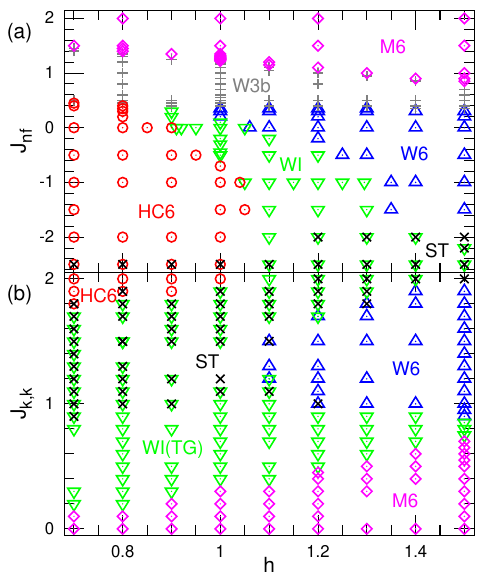}
\caption{
Dynamic phase diagram for $L=128$.
(a) $h$ vs. $J_{\mathrm{nf}}$  at $J_{k,k}=2$ ($J_{k,[k+2]}= J_{k,[k+3]}= J_{\mathrm{nf}}$).
(b) $h$ vs. $J_{k,k}$ at $J_{\mathrm{nf}}=-2$.
The red circles, gray pulses, magenta diamonds, black crosses,
blue up-pointing triangles, and green down-pointing triangles represent
the HC6, W3b, M6, ST, W6, and WI modes, respectively.
The overlapped symbols represent the coexistence of two modes owing to hysteresis.
The WI modes in (b) exhibit TG waves.
}
\label{fig:pdnf}
\end{figure}

\begin{figure}[tbh]
\includegraphics[]{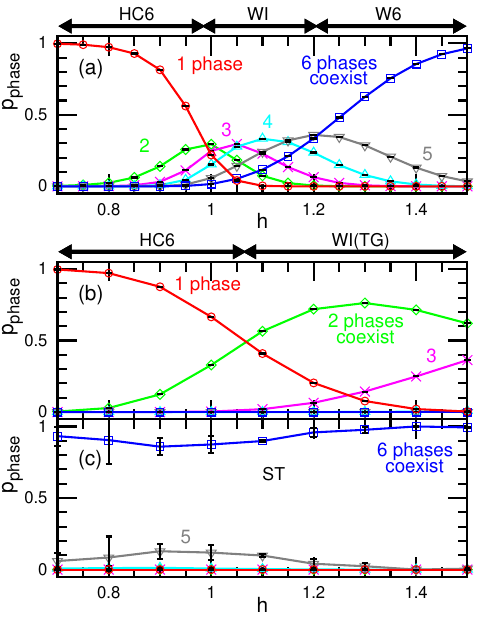}
\caption{
Probabilities $p_{\mathrm{phase}}$ of one-phase and multi-phase coexistence states as functions of $h$
for (a) $J_{\mathrm{nf}}=-0.5$ and (b),(c) $J_{\mathrm{nf}}=-2.5$ at $J_{k,k}=2$ and $L=128$.
The bidirectional arrows at the top of (a) and (b) represent the ranges of dynamic modes.
At $J_{\mathrm{nf}}=-2.5$, the ST mode coexists with the other modes [HC6 or WI(TG)] via hysteresis.
}
\label{fig:md05}
\end{figure}

In this study, we simulated the six-state Potts models in two-dimensional space using Monte Carlo (MC) method.
For square and hexagonal lattices, square and rhombus simulation cells with a side length of $L$ are employed, respectively, under periodic boundary conditions. 
The total number of sites is $N=L^2$.
Snapshots of the hexagonal lattices are displayed in a rectangular cell of $L \times \sqrt{3}L/2$.
We mainly simulated the dynamics under cyclically symmetric conditions (see Fig.~S1).
The states flip only into the neighboring states ($s=k \to [k\pm 1]$), and the flipping energy is uniform as $h_{k,[k+1]}=h$.
The contact energy $J_{k,k'}$ depends on the distance $j$ along the state flips, where $[k'=k+ j]$ and $j\in [0,3]$, as shown in Fig.~S1(b).
Three contact energy parameters, $J_{k,k}$, $J_{k,[k+2]}$, and $J_{k,[k+3]}$, are varied, while keeping $J_{k,[k+1]}=0$ as the reference energy level.
In Sec.~\ref{sec:w3}, 
we also discuss the effects of the asymmetry contact energy $J_{k,[k+2]}$ for even- and odd-numbered states
for the SP wave modes comprising either three states.

In the MC simulations,
a site is randomly selected, and
its flip to a neighboring state is performed using the Metropolis MC algorithm,  with the acceptance ratio calculated as $p_{s_is'_i}=\min[1, \exp(-\Delta H_{s_is'_i})]$.
This flipping is attempted $N$ times per MC step (time unit).
We previously verified that the choice between the Metropolis and Glauber schemes for the MC update causes only a minor difference.\cite{nogu24a,nogu26b}
The statistical errors are calculated from three or more independent runs
(ten independent runs are used for the coarsening dynamics).
The cluster size $n_{\mathrm{cl}}$ is calculated as the average across the clusters of all states.
The neighboring sites occupied by the same state belong to the same cluster.
We present the results obtained using square lattices in Secs.~\ref{sec:rev}--\ref{sec:w3}, 
and compare them with those using hexagonal lattices in Sec.~\ref{sec:hex}.

\section{Results and Discussion}\label{sec:results}

\subsection{Dynamic Modes}\label{sec:rev}

Since six can be factorized into two and three, several modes with factorized symmetry emerge,\cite{nogu25b} 
as in equilibrium systems.\cite{ditz80,gres81,plas86,bena88,taka20}
Figure~\ref{fig:pdnf}(a) shows
the dynamic phase diagram for $h$ vs. $J_{\mathrm{nf}}$ at $J_{k,k}=2$, where $J_{\mathrm{nf}}$ is the energy for nonflip contacts as $J_{k,[k+2]}=J_{k,[k+3]}=J_{\mathrm{nf}}$.
Since the modes at $J_{\mathrm{nf}}\ge -1.5$ were reported in our previous paper,\cite{nogu25b}
we briefly explain them here.

At $-1.5\lesssim J_{\mathrm{nf}}\lesssim 0$, 
the following three modes emerge from low to high flipping energies $h$: the homogeneous cycling mode (HC), intermediate wave mode (WI), and wave mode comprising six states (W6) [see the middle of Fig.~\ref{fig:pdnf}(a)].
In the HC mode, the entire lattices are covered by a single state for most periods, similar to an order phase in equilibrium.
However, the dominant states cyclically change through nucleation and growth, as follows: $s=0\to 1 \to 2 \to 3\to 4 \to 5\to 0$ [see Fig.~S2(e)].
In the W6 mode at high $h$, all states form domains in which the boundaries of the $s=k$ to $s=[k+1]$ domains move in the $s=k$-domain directions [see Fig.~S2(a)]. 
In the WI mode at intermediate $h$, multiple states, although not all, form domains for most periods [states with $N_S/N\simeq 0$ exist for most periods, as shown in Fig.~S2(c)].
To distinguish between these modes, we calculated the probabilities $p_{\mathrm{phase}}$ of the one-phase and multi-phase coexistence states. 
It is considered that $n$ phases spatially coexist when $n$ states satisfy $N_s/N>0.05$,
and only a single phase dominantly exists when $N_s/N>0.95$, where $s \in [0,5]$.
As $h$ increases,
the fraction of the single-phase dominance decreases, and that of six-phase coexistence increases, respectively [see Fig.~\ref{fig:md05}(a)].
We considered that the HC and W6 modes emerge when the single- and six-phase states have the largest $p_{\mathrm{phase}}$, respectively.
When an intermediate number of phases (two to five) is the largest, it is in the WI mode [see the bidirectional arrows at the top of Fig.~\ref{fig:md05}(a)]. 
The details of the W6 and WI modes are discussed  in  Sec.~\ref{sec:w6}.

\begin{figure}[tbh]
\includegraphics[]{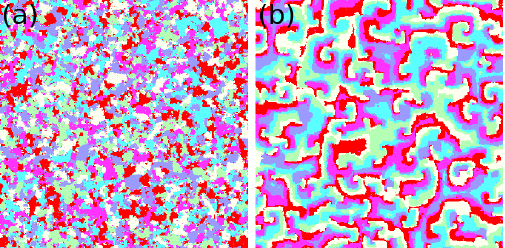}
\includegraphics[]{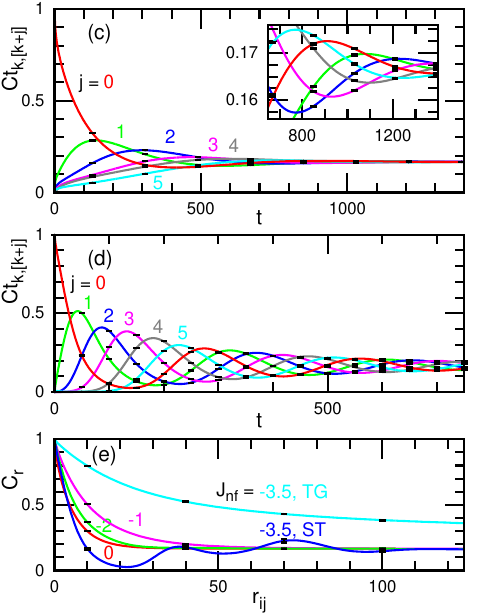}
\caption{
Wave forms in the W6 mode for $h=1.5$, $J_{k,k}= 2$, and $L=512$.
(a) Snapshot of DS waves at $J_{\mathrm{nf}}=0$.
(b) Snapshot of SP waves at $J_{\mathrm{nf}}=-2$.
The light yellow, light green, cyan, blue, magenta, and red sites (light to dark in grayscale) 
represent $s=0$, $1$, $2$, $3$, $4$, and $5$, respectively.
(c),(d) Time correlation $Ct_{k,[k+j]}(t)$ between the $s=k$ and $[k+j]$ states
at  (c)  $J_{\mathrm{nf}}=0$ and (d) $J_{\mathrm{nf}}=-2$.
The inset of (c) shows the magnified graph around the first peak at $j=0$.
(e) Spatial autocorrelation $C_{\mathrm{r}}(r)$
at $J_{\mathrm{nf}}=0$, $-1$, $-2$, and $-3.5$.
At $J_{\mathrm{nf}}=-3.5$,
the TG and ST waves are formed via hysteresis.
The error bars are shown at several data points, but most of them are smaller than the line thickness.
}
\label{fig:w6a}
\end{figure}

At $J_{\mathrm{nf}} \simeq 1$ and $J_{\mathrm{nf}} \simeq 2$, 
SP waves comprising three (even- or odd-numbered) states and a mixture of six states emerge, 
which are referred to as the W3b and M6 modes, respectively [see the upper region of Fig.~\ref{fig:pdnf}(a)].
SP waves comprising three states also emerge for $J_{k,[k+3]}\lesssim -1$ at $J_{k,k}=2$ and $J_{k,[k+2]}=0$ (see Fig.~S3).
Although previously considered qualitatively identical,\cite{nogu25b}
these waves are classified as backward and forward waves (W3b and W3f), as described in  Sec.~\ref{sec:w3}.

At $J_{\mathrm{nf}}\lesssim -1.5$, ST waves can emerge in addition to the HC6 and WI modes [see the bottom region of Fig.~\ref{fig:pdnf}(a)].
When two observed modes do not switch or switch only too infrequently to calculate an average, the symbols of both modes are indicated in the phase diagram.
The details of this region are discussed  in  Secs.~\ref{sec:tg} and \ref{sec:cd}.
The $J_{k,k}$ dependence are discussed  in Sec.~\ref{sec:mix}.

\subsection{Waves of six states}\label{sec:w6}

\subsubsection{Disordered (DS) and Spiral (SP) Waves}\label{sec:sp}

In the W6 mode,
 DS and SP waves are formed
by the contact interactions of the standard Potts model ($J_{\mathrm{nf}}=0$) and 
 repulsive interactions for nonflip contacts ($J_{\mathrm{nf}}\simeq -2$), respectively.
In the DS waves, the domains do not have a spiral shape,
since the boundaries of nonflip contacts ($s=k$ and $[k+2]$ or $[k+3]$)
do not move ballistically,  as shown in Fig.~\ref{fig:w6a}(a) and Movie~S1.
In contrast, SP waves have significantly shorter nonflip-contact boundaries owing to their high contact energies
(see Fig.~\ref{fig:w6a}(b) and the second half of Movie~S2).
Previously, these two wave modes have been distinguished only from visual images.
Here, we discuss the quantitative difference using correlation functions.

\begin{figure}[tbh]
\includegraphics[]{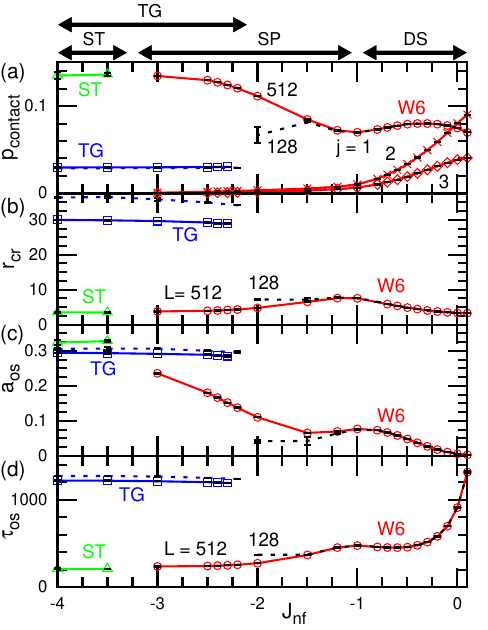}
\caption{
Dependence on the nonflip contact energy $J_{\mathrm{nf}}$ at $h=1.5$ and $J_{k,k}= 2$.
(a) Contact probabilities $p_{\mathrm{contact}}$.
(b) Correlation length $r_{\mathrm{cr}}$.
(c),(d) Amplitude $a_{\mathrm{os}}$ and time $\tau_{\mathrm{os}}$ 
of the first peak in the time correlation $Ct_{k,k}(t)$.
The solid and dashed lines represent the data for $L=512$ and $128$, respectively.
The green and blue lines represent the data of ST and TG waves, respectively.
The data for the ST waves at $L=128$ is not shown, since it is strongly restricted by the system size owing to the wavelength $\sim L$.
The bidirectional arrows at the top represent the ranges of the dynamic modes at $L=512$.
}
\label{fig:m15crt}
\end{figure}

The time correlation between the $s=k$ and $s=[k+j]$ states at each site is averaged for all sites over sufficiently long time periods as 
$Ct_{k,[k+j]}(t)= \langle \delta_{s_i(t_0),[s_i(t-t_0)-j]}\rangle$,
where $\langle ... \rangle$ denotes the average.
Figures~\ref{fig:w6a}(c) and (d) show  $Ct_{k,[k+j]}(t)$ for the DS and SP waves at $L=512$, respectively.
In both cases, $Ct_{k,[k+j]}(t)$ exhibits oscillatory decay, and the peaks sequentially appear along the cycle ($j= 1 \to 2 \to 3\to 4 \to 5\to 0$). The oscillation amplitude $a_{\mathrm{os}}$  and  period $\tau_{\mathrm{os}}$ are determined from the first peak of the autocorrelation function ($Ct_{k,k}(t)$):
$\tau_{\mathrm{os}}$ is the time when the first peak occurs, and 
$a_{\mathrm{os}}=a_{\mathrm{pk}}-\rho_{\mathrm{sat}}$, where $a_{\mathrm{pk}}$ is the peak amplitude, and $\rho_{\mathrm{sat}}$ is the saturation density, i.e., the oscillation center ($\rho_{\mathrm{sat}}=1/6$ for the W6 and M6 modes).
As $J_{\mathrm{nf}}$ decreases, $a_{\mathrm{os}}$ increases, and $\tau_{\mathrm{os}}$ decreases, representing  overall trends in the W6 mode; however, they exhibit shallow peaks at $J_{\mathrm{nf}}\simeq -1$ [see Fig.~\ref{fig:m15crt}(c) and (d)].

\begin{figure}[tbh]
\includegraphics[]{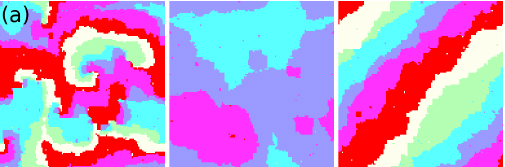}
\includegraphics[]{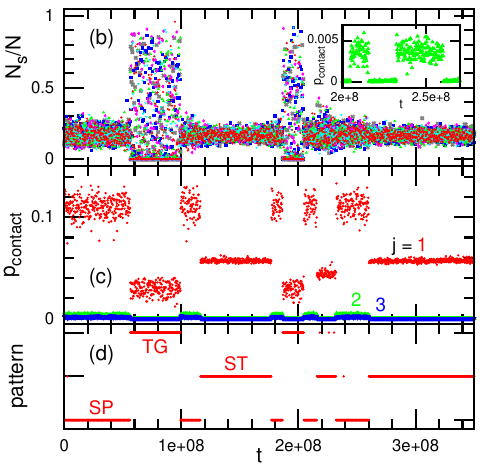}
\caption{
Temporal coexistence of SP, ST, and TG waves at $h=1.5$, $J_{\mathrm{nf}}=-2$, $J_{k,k}=2$, and $L=128$.
(a) Sequential snapshots (SP, TG, and ST waves at $t=800~000$, $1~040~000$, and $1~580~000$ from left to right).
(b)--(d) Time evolution of the number densities of each state ($N_s/N$), instantaneous contact probabilities ($p_{\mathrm{contact}}$), and wave patterns.
The gray, green, light blue, blue, cyan, and red symbols in (b) represent the number densities of $s=0$, $1$, $2$, $3$, $4$, and $5$, respectively.
The red, green, and blue symbols in (c) represent $p_{\mathrm{contact}}$
for the contact between $s=k$ and $[k+j]$ states with $j=1$, $2$, and $3$, respectively.
The inset of (b) shows a magnified plot at $j=2$ for $2\times 10^8 < t < 2.7 \times 10^8$.
(d) The dots in the upper, middle, and bottom regions represent TG, ST, and SP waves, respectively.
}
\label{fig:m15d2t}
\end{figure}

The spatial autocorrelation function $C_{\mathrm{r}}({\mathbf{r}}) =  \langle\delta_{s_i,s_j}\delta({\mathbf{r}(t)}_{i,j}-{\mathbf{r}(t)}) \rangle$ is 
calculated along the $x$ and $y$ axes [$(x,y)=(1, 0)$ and $(0, 1)$].
In the W6 mode, $C_{\mathrm{r}}({\mathbf{r}})$ exhibits monotonic decay [see the curves for $J_{\mathrm{nf}}= 0$, $-1$, and $-2$ in Fig.~\ref{fig:w6a}(e)].
The correlation length $r_\mathrm{cr}$ is defined as
the half-relaxation length: $C_{\mathrm{r}}(r_{\mathrm{cr}})=(1+ \rho_{\mathrm{sat}})/2$.
The length $r_\mathrm{cr}$ also exhibits a shallow peak at $J_{\mathrm{nf}}\simeq -1$ [see Fig.~\ref{fig:m15crt}(b)].

Figure~\ref{fig:m15crt}(a) shows the contact probability $p_{\mathrm{contact}}$ between different states on neighboring sites. When all the neighbors of $s=k$ sites are occupied by the $s=[k+ 1]$ states (alternative arrangement in the checkerboard pattern), $p_{\mathrm{contact}}=1$, $0$, and $0$ for the flip distances $j=1$, $2$, and $3$, respectively.
The probabilities $p_{\mathrm{contact}}$ for nonflip contacts ($j=2$ and $3$) decrease with decreasing $J_{\mathrm{nf}}$. In contrast, $p_{\mathrm{contact}}$ of the flip contact ($j=1$) exhibits a minimum at $J_{\mathrm{nf}}\simeq -1$ [see the three red curves in Fig.~\ref{fig:m15crt}(a)].
Thus, these four quantities ($a_{\mathrm{os}}$, $\tau_{\mathrm{os}}$, $r_\mathrm{cr}$, and $p_{\mathrm{contact}}$) have a maximum or minimum at $J_{\mathrm{nf}}\simeq -1$.
This phase boundary matches the results of visual inspection.
Therefore, we conclude that the dynamic mode changes from DS to SP waves at $J_{\mathrm{nf}}\simeq -1$ for $h=1.5$.
However, we do not separate them in the phase diagram [Fig.~\ref{fig:pdnf}(a)],
since these maxima and minima are uncertain for lower $h$ due to the temporal coexistence of the WI mode.

\begin{figure}[tbh]
\includegraphics[]{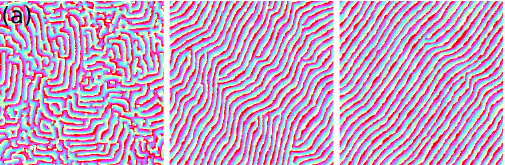}
\includegraphics[]{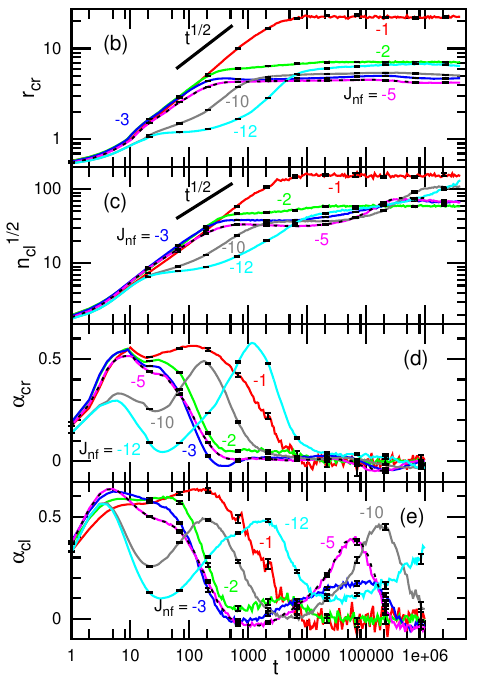}
\caption{
Coarsening dynamics at $h=1$ and $J_{k,k}=2$.
(a) Sequential snapshots at $J_{\mathrm{nf}}=-5$ and $L=1024$ ($t=100~000$, $500~000$, and $4~000~000$ from left to right). 
(b)--(e) Time evolution of 
(b) the correlation length $r_{\mathrm{cr1}}$, (c) root of the mean cluster size ${n_{\mathrm{cl}}}^{1/2}$, 
(d) coarsening exponent $\alpha_{\mathrm{cr}}$ of $r_{\mathrm{cr1}}$, and (e) coarsening exponent $\alpha_{\mathrm{cl}}$ of ${n_{\mathrm{cl}}}^{1/2}$.
The solid lines represent the data for $J_{\mathrm{nf}}=-1$, $-2$, $-3$, $-5$, $-10$, and $-12$ at $L=1024$.
The black dashed lines represent the data for $J_{\mathrm{nf}}=-5$ at $L=2048$.
The short black lines in (b) and (c) represent the slope of $t^{1/2}$.
The error bars are shown at several data points.
}
\label{fig:sd}
\end{figure}

\subsubsection{Target (TG) and Stripe (ST) Waves}\label{sec:tg}

Under very strong repulsion at the nonflip contacts ($J_{\mathrm{nf}}\lesssim -3$) for $h=1.5$,
the SP waves are suppressed, since nonflip contacts occur at the centers of spiral shapes. When SP waves or randomly mixed states are used as initial states,
ST waves are formed [see Movie~S3 and the right snapshots in Figs.~\ref{fig:m15d2t}(a) and \ref{fig:sd}(a)]. 
In contrast, TG waves are formed from a uniform initial state;
an $s=[k+1]$ domain appears nested inside of an $s=k$ domain [see the middle snapshot in Fig.~\ref{fig:m15d2t}(a)]. 
The centers of the TG waves randomly appear through nucleation and growth,
unlike TG waves in experiments of the Belousov--Zhabotinsky reaction, 
in which the TG centers are typically fixed by pacemakers.\cite{mikh06}
These two modes do not switch at $J_{\mathrm{nf}}<-3.5$.
Since the TG wavelength and period are determined by the nucleation time,
they are greater than those of the SP waves, whereas those of the ST and SP waves are comparable [see Fig.~\ref{fig:m15crt}(b) and (d)]. 
The spatial correlation, $C_{\mathrm{r}}(r)$,  exhibits an oscillation in the ST waves, whereas it monotonically decays in the TG waves
[see Fig.~\ref{fig:w6a}(e)].

At an intermediate repulsion, three wave modes (TG, ST, and SP) temporally coexist
in a small system, as shown in Fig.~\ref{fig:m15d2t} and Movie~S4 for $J_{\mathrm{nf}}=-2$ and $L=128$.
We distinguish these modes using the number densities $N_s/N$ of the states and instantaneous contact probabilities $p_{\mathrm{contact}}$ [Fig.~\ref{fig:m15d2t}(b) and (c)].
Since $p_{\mathrm{contact}}$ for $j=2$ is higher in the SP waves than in the TG and ST waves,
it is considered that the SP waves are formed when $p_{\mathrm{contact}}>p_{\mathrm{th}}$,
where $p_{\mathrm{th}}$ is determined from the time evolution plots for each condition
[$p_{\mathrm{th}}=0.001$ for $J_{\mathrm{nf}}=-2$ and $h=1.5$, as shown in the inset of Fig.~\ref{fig:m15d2t}(b)].
The ST and TG waves are distinguished based on $N_s/N$: 
ST waves are formed when  $N_s/N>0.05$ for all states, whereas TG waves are formed otherwise [see Fig.~\ref{fig:m15d2t}(d)].
For larger system sizes, the stable range of SP waves is broader [compare the red solid and black dashed curves in Fig.~\ref{fig:m15crt}]; for $L=512$ and $J_{\mathrm{nf}}=-2$, the SP mode does not switch to the TG mode, whereas the other transition occurs (see Movie~S2).
This is because the SP-to-TG transition becomes exponentially rare with increasing $L$,
whereas the opposite transition occurs more frequently  in larger $L$.
This is similar to the transition between the HC and SP modes reported in Ref.~\onlinecite{nogu24a}.
A similar exponential-size dependence has been reported for transient spiral chaos in excitable media.\cite{qu06,sugi15}

The TG waves are categorized into the WI mode based on the ratios of $p_{\mathrm{phase}}$.
As $J_{\mathrm{nf}}$ decreases at $h\simeq 1.2$, $p_{\mathrm{contact}}$ for nonflip contacts ($j=2$ and $3$)
exponentially decrease [see Fig.~S4].
At $J_{\mathrm{nf}}\simeq -0.5$, the waves do not exhibit a TG shape, and  a significant number of nonflip contacts exist [see Fig.~S2(d)]. 
Conversely, at $J_{\mathrm{nf}}\simeq -1.5$, clear TG waves are formed, and nonflip contacts rarely appear; even if a nonflip contact appears,
it quickly disappears instead of growing into an SP wave [see Fig.~S5].
Thus, the boundary of the TG waves is at $J_{\mathrm{nf}}\simeq -1$ for $h\simeq 1.2$.
As $h$ decreases at $J_{\mathrm{nf}}=-2.5$, the number of coexisting states decreases,
and the HC6 mode appears at $h\leq 1$ owing to an increase in the nucleation period;
these modes coexist with the ST waves via hysteresis [see Fig.~\ref{fig:md05}(b) and (c)].

\begin{figure}[tbh]
\includegraphics[]{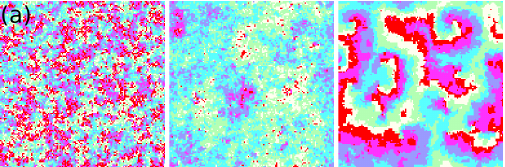}
\includegraphics[]{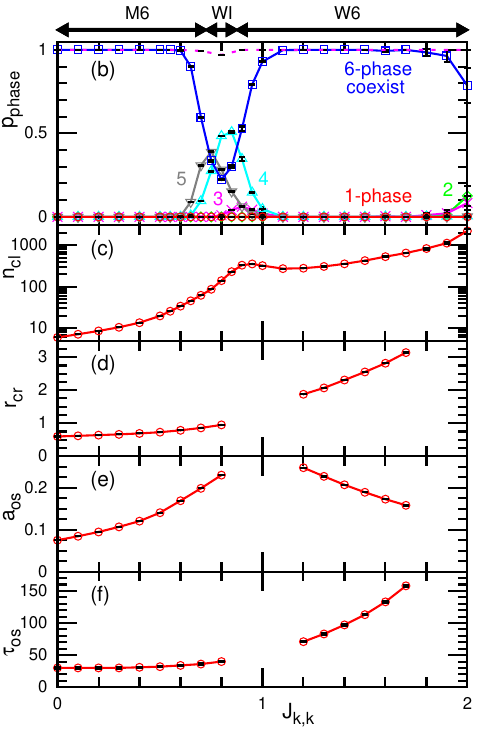}
\caption{
Dependence on the contact energy $J_{k,k}$ at $h=1.5$ and $J_{\mathrm{nf}}= -2$.
(a) Snapshots at $J_{k,k}=0$, $0.8$, and $1.5$ (from left to right).
(b) Probabilities $p_{\mathrm{phase}}$ of one-phase and multi-phase coexistence states.
The solid lines represent $p_{\mathrm{phase}}$ at $L=128$.
The magenta dashed line represents $p_{\mathrm{phase}}$ of the six-phase coexistence state
at  $L=256$, whereas the other states at $L=256$ are omitted for clarity.
The bidirectional arrows at the top represent the ranges of dynamic modes at $L=128$.
(c)--(f) Dependence of (c) the mean cluster size $n_{\mathrm{cl}}$, (d) correlation length $r_{\mathrm{cr}}$, and
(e) amplitude $a_{\mathrm{os}}$ and (f) time $\tau_{\mathrm{os}}$ 
of the first peak in the time correlation $Ct_{k,k}(t)$ at $L=128$.
At $J_{k,k}\simeq 1$, $r_{\mathrm{cr}}$, $a_{\mathrm{os}}$, and $\tau_{\mathrm{os}}$ are not shown,
since different modes temporally coexist.
}
\label{fig:m15d2e}
\end{figure}

\subsubsection{Coarsening Dynamics}\label{sec:cd}

In the Potts models,
coarsening into an equilibrium state follows the LAC law ($\ell \propto t^{1/2}$).\cite{nogu26b,bray94,gres88}
Accordingly, for $h=0$, both the correlation length $r_{\mathrm{cr}}$ and cluster length ${n_{\mathrm{cl}}}^{1/2}$ (the root of the cluster size)
are proportional to $t^{1/2}$  (see the red lines in Fig.~S6).
For $h\ne 0$, these lengths saturate into the characteristic lengths of the steady wave patterns.
Interestingly, these lengths more rapidly increases slightly before saturation in the W6 modes [see Fig.~S6(b) and (c)].\cite{nogu26b}
These temporal speedups are clearly captured by the time evolution of the coarsening exponents
$\alpha_{\mathrm{cr}}$ and $\alpha_{\mathrm{cl}}$ [see Fig.~S6(d) and (e)].
These coarsening exponents are calculated as\cite{nogu26b,huse86}
\begin{eqnarray}
  \alpha_{\mathrm{cr}}(t) &=& \log_b[r_{\mathrm{cr}}(bt)/r_{\mathrm{cr}}(t)], \\
  \alpha_{\mathrm{cl}}(t) &=& (1/2)\log_b[n_{\mathrm{cl}}(bt)/n_{\mathrm{cl}}(t)],
\end{eqnarray}
with $b=4$.
The heights of the exponent peaks increase with increasing number of states $q$
and decrease with decreasing $J_{\mathrm{nf}}$ (DS to SP waves), as shown in Fig.~S7.\cite{nogu26b}

In coarsening under strong repulsion at nonflip contacts ($J_{\mathrm{nf}}\lesssim -3$),
SP waves are temporally formed, and subsequently, ST waves are formed through the fusion of crescent-shaped domains [see Fig.~\ref{fig:sd}(a)]. During 
this stripe-pattern formation, the domains become tangentially larger. This change can be captured by an increase in ${n_{\mathrm{cl}}}^{1/2}$, while maintaining the value of $r_{\mathrm{cr}}$ at $t\simeq 10^{5}$ [see Fig.~\ref{fig:sd}(b)--(e)]. In other words, the growth in the wave-propagation direction saturates faster than that in the perpendicular direction.
The correlation and cluster lengths are characteristic quantities for these two growth types.

\begin{figure}[tbh]
\includegraphics[]{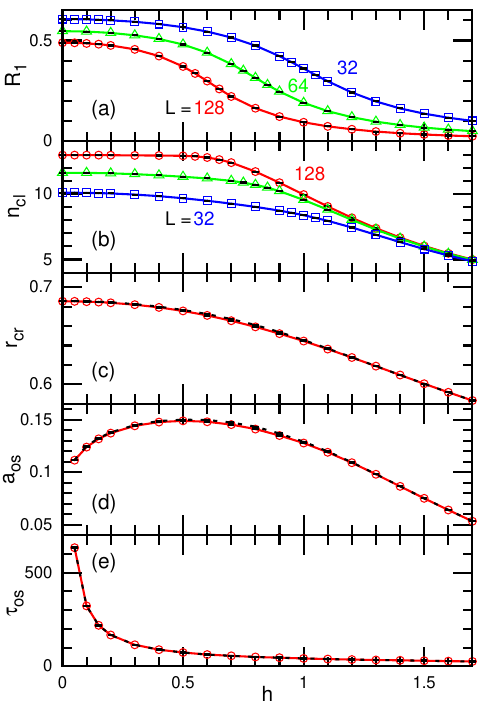}
\caption{
Dependence on the flip energy $h$ in the six-state mixing mode (M6) at $J_{k,k}= 0$ and  $J_{\mathrm{nf}}= -2$.
(a) Order parameter $R_1$. (b) Mean cluster size $n_{\mathrm{cl}}$. (c) Correlation length  $r_{\mathrm{cr}}$. 
(d),(e) Amplitude $a_{\mathrm{os}}$ and time $\tau_{\mathrm{os}}$ 
of the first peak in the time correlation $Ct_{k,k}(t)$.
The three lines in (a) and (b) represent the data at $L=32$, $64$, and $128$
[from top to bottom in (a) and from bottom to top in (b)].
The solid and dashed lines in (c)--(e) represent the data at $L=128$ and $32$, respectively. These lines overlap.
}
\label{fig:d2e0}
\end{figure}

\begin{figure}[tbh]
\includegraphics[]{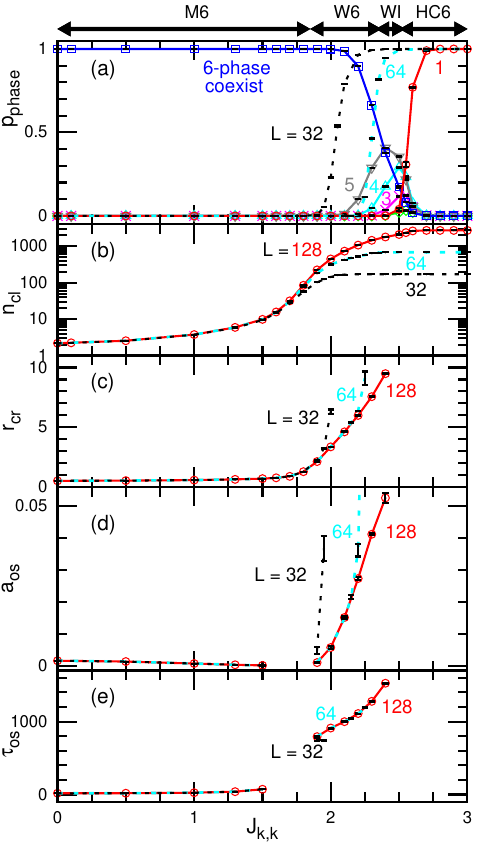}
\caption{
Dependence on the contact energy $J_{k,k}$ at $h=1.5$ and $J_{\mathrm{nf}}= 0$.
(a) Probabilities $p_{\mathrm{phase}}$ of one-phase and multi-phase coexistence states.
The solid lines represent $p_{\mathrm{phase}}$ at $L=128$.
The black and light blue dashed lines represent $p_{\mathrm{phase}}$ of the single-phase state
at  $L=32$ and $64$, respectively, whereas the other states at $L=32$ and $64$ are omitted for clarity.
The bidirectional arrows at the top represent the ranges of dynamic modes at $L=128$.
(b)--(e) Dependence of (b) the mean cluster size $n_{\mathrm{cl}}$, (c) correlation length $r_{\mathrm{cr}}$,
and (d) amplitude $a_{\mathrm{os}}$ and (e) time $\tau_{\mathrm{os}}$ 
of the first peak in the time correlation $Ct_{k,k}(t)$.
}
\label{fig:m15d0}
\end{figure}

\begin{figure}[tbh]
\includegraphics[]{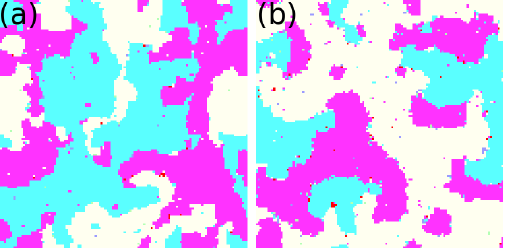}
\includegraphics[]{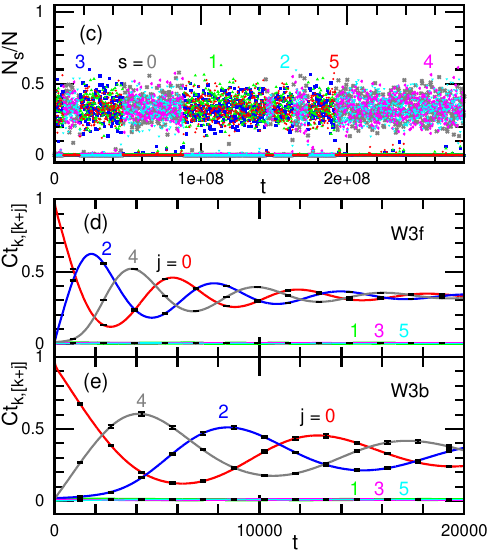}
\caption{
SP waves comprising three even- or odd-numbered states at $h=1$, $J_{k,k}= 2$, and $L=128$.
(a),(c),(d) Forward waves (W3f) at $J_{k,[k+2]}= 0$ and $J_{k,[k+3]}= -1$.
(b),(e) Backward waves (W3b) at $J_{k,[k+2]}= J_{k,[k+3]}= 0.5$.
(a),(b) Snapshots.
The light yellow,  cyan, and magenta sites
represent $s=0$, $2$, and $4$, respectively.
A small number of odd-numbered ($s=1$, $3$, and $5$) states exist at domain boundaries.
(c) Time evolution of the number densities of each state.
(d),(e) Time correlation between $k$ and $[k+j]$ states.
The error bars are shown at several data points.
}
\label{fig:w3a}
\end{figure}

\subsubsection{Transitions from Wave to Mixing Modes}\label{sec:mix}
As the attraction $J_{k,k}$ between the same states decreases at $J_{\mathrm{nf}}= -2$, 
the domains become smaller, their boundaries fluctuate more widely,
and subsequently, all states are mixed as small clusters (M6) [see Fig.~\ref{fig:m15d2e}(a)].
Since nonflip contacts are still avoided, the TG waves are formed in an intermediate range of $J_{k,k}$
[see Figs.~\ref{fig:pdnf}(b) and \ref{fig:m15d2e}(a)].
Since all states spatially coexist in both the W6 and M6 modes,
they cannot be distinguished through $p_{\mathrm{phase}}$ [see Fig.~\ref{fig:m15d2e}(b)].
However, the time correlation amplitude $a_{\mathrm{os}}$ exhibits a different dependence on $J_{k,k}$;
$a_{\mathrm{os}}$ increases as $J_{k,k}$ approaches unity [see Fig.~\ref{fig:m15d2e}(e)].
The other quantities, $n_{\mathrm{cl}}$, $r_{\mathrm{cl}}$, and $\tau_{\mathrm{os}}$, also exhibit slight differences in their slopes with respect to $J_{k,k}$  [see Fig.~\ref{fig:m15d2e}(c), (d), and (f)].
Note that $r_{\mathrm{cl}}$, $a_{\mathrm{os}}$, and $\tau_{\mathrm{os}}$ are not calculated  at $J_{k,k}\simeq 1$,
since different modes temporally  coexist there.

Next, we investigate the properties of the M6 mode.
As $h$ decreases at $J_{k,k}=0$ [bottom of the phase diagram shown in Fig.~\ref{fig:pdnf}(b)], 
the clusters become larger, whereas the correlation length $r_{\mathrm{cr}}$ becomes slightly longer [see Fig.~\ref{fig:d2e0}(b) and (c)].
The order parameter $R_1$ for the occupation by a single state has a sigmoidal shape owing to the change in the cluster size [see Fig.~\ref{fig:d2e0}(a)]:
\begin{equation}
  R_1 = \frac{1}{N}\bigg\langle \bigg|\sum_j^N \exp\Big(\frac{2\pi{\mathrm{i}}s_j}{6}\Big)\bigg| \bigg\rangle.
\end{equation}
However, the inflection point shifts to a lower $h$ with increasing system size,
so that the sigmoidal shape is caused by a finite-size effect and disappears with $L\to \infty$.
As $h\to 0$, the system approaches an equilibrium state; thus, $a_{\mathrm{os}}$ vanishes, and $\tau_{\mathrm{os}}$ diverges.
The correlations ($r_{\mathrm{cr}}$, $a_{\mathrm{os}}$, and $\tau_{\mathrm{os}}$)  have negligible finite-size dependence [see Fig.~\ref{fig:d2e0}(c)--(e)].

Under the standard Potts interaction ($J_{\mathrm{nf}}=0$),
the system exhibits a direct M6--W6 transition with increasing $J_{k,k}$.
The oscillation in the time autocorrelation $Ct_{k,k}(t)$ vanishes at the transition point ($a_{\mathrm{os}} \to 0$),
and the slopes of $n_{\mathrm{cl}}$ and $r_{\mathrm{cr}}$ change [see Fig.~\ref{fig:m15d0}].
System-size dependence is obtained in the range of the W6 mode, but not in the M6 range.
Thus, the M6 and W6 modes can be distinguished based on their time correlations.

\begin{figure}[tbh]
\includegraphics[]{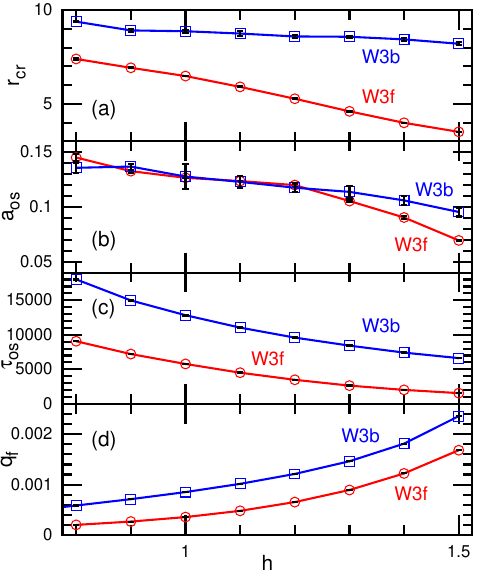}
\caption{
Dependence on the flip energy $h$ in the W3f and W3b modes at $J_{k,k}= 2$ and $L=128$.
(a) Correlation length $r_{\mathrm{cr}}$. (b),(c) Amplitude $a_{\mathrm{os}}$ and time $\tau_{\mathrm{os}}$
of the first peak in the time correlation $Ct_{k,k}(t)$.
(d) Flow rate $q_{\mathrm{f}}$ between successive states.
The red circles and blue squares represent the data for W3f and W3b
at $J_{k,[k+2]}= 0$ and $J_{k,[k+3]}= -1$ and at $J_{k,[k+2]}= J_{k,[k+3]}= 0.5$, respectively.
}
\label{fig:w3crt}
\end{figure}

\begin{figure}[tbh]
\includegraphics[]{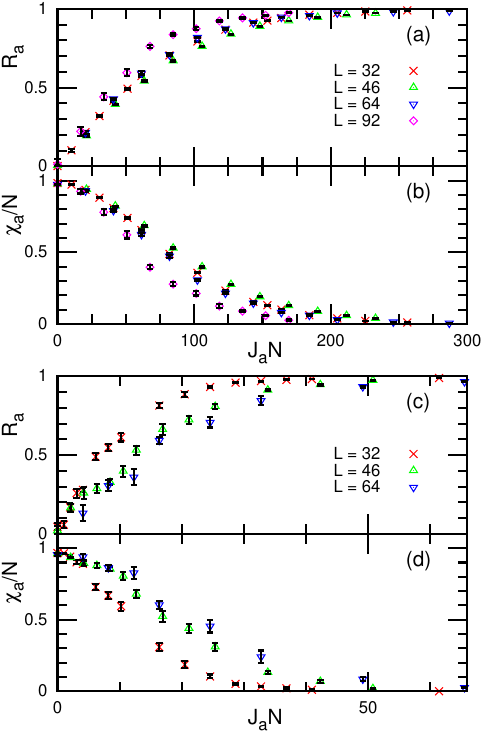}
\caption{
Ratio of SP waves comprising even- and odd-numbered states
 in (a),(b) W3f and (c),(d) W3b at $h=1$ and $J_{k,k}=2$.
The order parameter $R_{\mathrm{a}}=\langle N_{\mathrm{even}}-N_{\mathrm{odd}} \rangle/N$ and susceptibility $\chi_{\mathrm{a}}$
are shown in (a),(c) and (b),(d), respectively, with the normalized asymmetric contact energy $J_{\mathrm{a}}N$.
In the W3f mode, $J_{k,[k+2]}= J_{\mathrm{a}}$ for $k=1,3,5$ 
and $J_{k,[k+2]}= -J_{\mathrm{a}}$ for $k=0,2,4$ at $J_{k,[k+3]}=-1$ and $L=32$, $46$, 
$64$, and $92$.
In the W3b mode, $J_{k,[k+2]}=0.6+ J_{\mathrm{a}}$ for $k=1,3,5$ 
and $J_{k,[k+2]}=0.6- J_{\mathrm{a}}$ for $k=0,2,4$ at $J_{k,[k+3]}=0.6$
and $L=32$, $46$, and $64$.
}
\label{fig:as}
\end{figure}

\subsection{Waves of Three States}\label{sec:w3}

SP waves comprising three states are formed 
for $J_{k,[k+3]}\lesssim -1$ and $J_{k,[k+2]}=0$, and for  $J_{\mathrm{nf}} \simeq 1$ at $J_{k,k}=2$ [see Fig.~\ref{fig:w3a}(a) and (b)].
Three even- or odd-numbered states form domains ($s=0$, $2$, and $4$, or $s=1$, $3$, and $5$, respectively).
The other three states exist to a small degree at domain boundaries
and occasionally nucleate into small domains, but they most eventually disappear [see Movies S5 and S6]. 
However, when the domains of the other three states grow over the entire lattices,
the three dominant states switch, as shown in Fig.~\ref{fig:w3a}(c).
These stochastic switches occur more frequently at lower $h$, reflecting lower densities of spiral centers.

In closely looking at Fig.~\ref{fig:w3a}(a) and (b), one can recognize that the directions of spirals are different [magenta ($s=4$) domains are inside of the cyan ($s=2$) domains in (a) but outside in (b)]. In Fig.~\ref{fig:w3a}(a) and Movie S5, the spirals wind forward ($s=0 \to 2 \to 4 \to 0$ or $s=1 \to 3 \to 5 \to 1$). Conversely, they wind backward ($s=0 \to 4 \to 2 \to 0$ or $s=1 \to 5 \to 3 \to 1$) in Fig.~\ref{fig:w3a}(b) and Movie S6.
Hence, we refer to these as forward waves (W3f) and backward waves (W3b), respectively.
To clarify this difference, we calculated the time correlations $Ct_{k,[k+j]}(t)$.
The peaks of $Ct_{k,[k+j]}(t)$ appear in the sequence $j=2$, $4$, and $0$ in the W3f mode,
whereas they appear in the sequence $j=4$, $2$, and $0$ in the W3b mode [see Fig.~\ref{fig:w3a}(d) and (e)].
However, their dependencies on $h$ are similar: $r_{\mathrm{cl}}$, $a_{\mathrm{os}}$, and $\tau_{\mathrm{os}}$ decrease and flow rate $q_{\mathrm{f}}$ increases with increasing $h$ (see Fig.~\ref{fig:w3crt}).
Here, the averages are taken for the periods that maintain the waves of either even- or odd-numbered states,
ensuring that the spatial and time correlations $C_{\mathrm{r}}({\mathbf{r}})$ and $Ct_{k,k}(t)$ of the major three states 
saturate to $\rho_{\mathrm{sat}}=N_{\mathrm{maj}}/3(N_{\mathrm{maj}} + N_{\mathrm{min}})$,
where $N_{\mathrm{maj}}$ and $N_{\mathrm{min}}$  are the numbers of major and minor states ($N_{\mathrm{maj}}=\langle N_0+N_2+N_4 \rangle$ and $N_{\mathrm{min}}=\langle N_1+N_3+N_5\rangle$ for even-numbered waves), as shown in Fig.~S8. 
Since the minor states exist to a small degree, $\rho_{\mathrm{sat}}$ is slightly smaller than $1/3$ [$0.324< \rho_{\mathrm{sat}}< 0.333$ for the data shown in Fig.~\ref{fig:w3crt}(a) and (b)].
The flow rate $q_{\mathrm{f}}$ is the average difference of forward and backward flips between neighboring states. Note that the entropy production rate\cite{herp18,agra25} is $qhq_{\mathrm{f}}$ for the steady state of $q$-state cyclic Potts models, since $\langle\Delta H_{\mathrm{int}}\rangle=0$ in steady states
and each flip locally stratifies the detailed balance condition.

Let us consider the mechanisms underlying these two types of waves.
The W3f waves emerge with strong repulsion between the diagonal states ($J_{k,[k+3]}\lesssim -1$
for $J_{k,[k+2]}=J_{k,[k+1]}=0$, and $J_{k,k}=2$).
The contact between the diagonal states ($s=k$ and $[k+3]$) can be avoided
when the $s=k$, $[k+2]$, and $[k+4]$ states form domains,
and the states of the boundary sites change through two-step forward flips ($s=k \to [k+1] \to [k+2]$ 
at the boundary of the $s=k$ and $[k+2]$ domains).
Hence, the W3f waves maintain a low interaction energy $H_{\mathrm{int}}$.
Although the ST waves comprising six states can avoid the diagonal contacts, the boundary between the $s=k$ and $[k+1]$ domains
is destabilized by nucleation and  growth of the $s=[k+2]$ state on the $[k+1]$ side of the boundary.

The W3b waves emerge at medium attraction at nonflip contacts ($J_{\mathrm{nf}}\simeq 1$ for $J_{k,[k+1]}=0$ and $J_{k,k}=2$).
The boundary between the $s=k$ and $[k+2]$ domains has a lower interfacial energy than that between the $s=k$ and $[k+1]$ domains, owing to the relatively higher repulsion between the $s=k$ and $[k+1]$ states. Hence, the W3 domains are stabilized.
The boundaries between the $s=k$ and $[k+2]$ domains move backward through four-step forward flips ($s=[k+2] \to [k+3] \to [k+4] \to [k+5] \to k$) at the boundaries.
The intermediate states in the four-step flips have lower energies than those in the two-step flips ($s=k \to [k+1] \to [k+2]$). 
For example, at a straight boundary between the $s=k$ and $[k+2]$ domains,
a flip to the $s=[k+1]$ state requires a high energy of $3J_{k,k}+J_{\mathrm{nf}} - 4J_{k,[k+1]}$.
In comparison, the three intermediate states in the four-step flips require
lower energies ($3J_{k,k}-2J_{\mathrm{nf}} - J_{k,[k+1]}$, $3J_{k,k}-3J_{\mathrm{nf}}$, and $3J_{k,k}-J_{\mathrm{nf}} - 3J_{k,[k+1]}$
for $s=[k+3]$, $[k+4]$, and $[k+5]$, respectively, when flips occur on the side of the $s=k$ domain). 
Thus, each site flips for two forward cycles per backward cycle ($s=k\to [k+4] \to [k+2]\to k$) of domains in the SP waves.
Although the boundaries of the diagonal domains ($s=k$ and $[k+3]$) have the same interfacial energy as those of the $s=k$ and $[k+2]$ domains, the diagonal boundary does not move ballistically because of symmetry; hence, it is driven out by  moving the $s=k$ and $[k+2]$ boundaries.
Therefore, the stability differences among the minor states produce forward and backward waves in the W3f and W3b modes.

Until here, the dynamics under cyclical symmetry have been described.
When the interactions for even- and odd-numbered states are asymmetric,
the ratio of waves comprising even- and odd-numbered states can take on unequal values in both the W3f and W3b modes.
We varied the contact energies $J_{k,[k+2]}$ of the even- and odd-numbered states asymmetrically, setting $J_{2n+1,[2n+3]} = J_0 + J_{\mathrm{a}}$
and $J_{2n,[2n+2]} = J_0 - J_{\mathrm{a}}$.
To quantify the wave ratio, we calculated 
the order parameter $R_{\mathrm{a}}=\langle u_{\mathrm{a}}\rangle$, where $u_{\mathrm{a}}=(N_{\mathrm{even}}-N_{\mathrm{odd}})/N$,  $N_{\mathrm{even}}=N_0+N_2+N_4$, and $N_{\mathrm{odd}}=N_1+N_3+N_5$.
Its susceptibility $\chi_{\mathrm{a}}$ is defined as $\chi_{\mathrm{a}} = N(\langle {u_{\mathrm{a}}}^2 \rangle - \langle u_{\mathrm{a}} \rangle^2)$. 
As $J_{\mathrm{a}}$ increases,  $R_{\mathrm{a}}$ changes sigmoidally from $-1$ to $1$, and the change becomes steeper at larger $N$. When $J_{\mathrm{a}}$ is normalized by $1/N$, the curves of $R_{\mathrm{a}}$ and  $\chi_{\mathrm{a}}/N$ for different $N$ overlap, as shown in Fig.~\ref{fig:as}.
Therefore, this change is a first-order transition, similar to that in the Ising model under a magnetic field.\cite{bind87,bind84}

\begin{figure}[]
\includegraphics[]{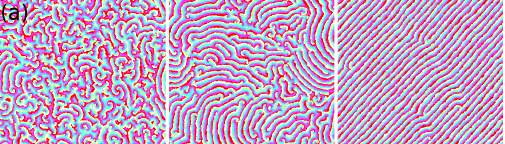}
\includegraphics[]{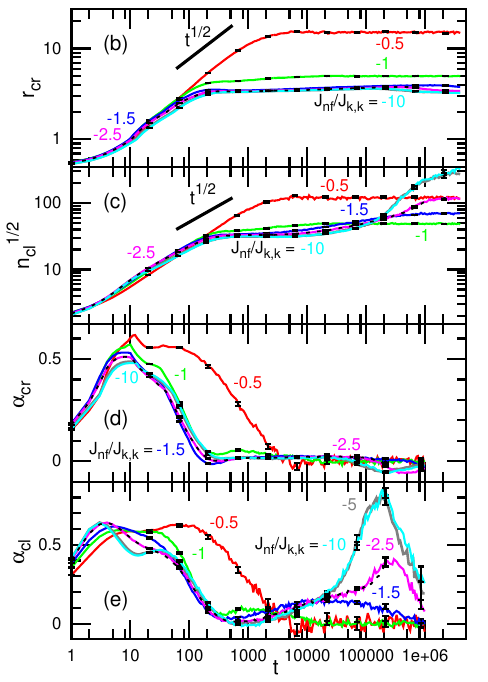}
\caption{
Coarsening dynamics in hexagonal lattices at $h=1$ and $J_{k,k}=1.2$.
(a) Sequential snapshots at $J_{\mathrm{nf}}/J_{k,k}=-2.5$ and $L=1024$ ($t=100~000$, $500~000$, and $3~000~000$ from left to right). 
(b)--(e) Time evolution of 
(b) the correlation length $r_{\mathrm{cr1}}$, (c) root of the mean cluster size ${n_{\mathrm{cl}}}^{1/2}$, 
(d) coarsening exponent $\alpha_{\mathrm{cr}}$ of $r_{\mathrm{cr1}}$, and (e) coarsening exponent $\alpha_{\mathrm{cl}}$ of ${n_{\mathrm{cl}}}^{1/2}$.
The solid lines represent the data for $J_{\mathrm{nf}}/J_{k,k}=-0.5$, $-1$, $-1.5$, $-2.5$, $-5$, and $-10$ at $L=1024$.
The black dashed lines represent the data for $J_{\mathrm{nf}}=-2.5$ at $L=2048$.
The short black lines in (b) and (c) represent the slope of $t^{1/2}$.
The error bars are shown at several data points.
}
\label{fig:sdtri}
\end{figure}

\subsection{Dynamic Modes in Hexagonal Lattices}\label{sec:hex}

In square lattices, the SP waves have rectangular shapes that elongate along the $x$ and $y$ axes,
reflecting the lattice grid.
When hexagonal lattices are employed, the effects of this lattice discreteness are significantly reduced.
The SP waves are more circular in the hexagonal lattices [see the left snapshot in Fig.~\ref{fig:sdtri}(a)].
However, we confirm that the dynamic modes reported in this paper and our previous paper\cite{nogu25b} also occur in hexagonal lattices
[see W6(DS), W6(SP), M6, W3b, TG, ST, W3f, M2W3, and M2HC3 in Fig.~S9(a)--(i), respectively].
In the M2W3 and M2HC3 modes, diagonal states ($s=k$ and $[k+3]$) form mixing phases owing to the attraction between diagonal states,
so that SP waves and homogeneous cycling of the three mixing phases occur, respectively (see Fig.~S3).

In our previous study,\cite{nogu26b} we confirmed that the coarsening dynamics from a random mixture to W6(SP) waves are independent of whether square or hexagonal lattices are used.
In this study, we examined the coarsening into the ST waves.
The ST waves are formed via intermediate SP waves, as in square lattices [see Fig.~\ref{fig:sdtri}(a)].
The correlation length $r_{\mathrm{cr}}$ saturates when SP waves are formed,
and the cluster length ${n_{\mathrm{cl}}}^{1/2}$ increases during the change from SP to ST waves [see Fig.~\ref{fig:sdtri}(b)--(e)].
The difference from those observed in the square lattices lies in the disappearance of the transient trap in the small clusters of ${n_{\mathrm{cl}}}^{1/2}\sim 10$ [compare $r_{\mathrm{cr}}$ and  ${n_{\mathrm{cl}}}^{1/2}$ at $100\lesssim t\lesssim 1000$
in Figs.~\ref{fig:sdtri} and \ref{fig:sd}]. This is due to the fewer neighbors in the square lattice. 
Previously, for the coarsening dynamics of Potts models at zero temperature,
more traps in metastable states were reported in square lattices\cite{olej13} than in hexagonal lattices.\cite{denh21}
The high $|J_{\mathrm{nf}}|$ condition for the ST waves produces effects similar to a strong quench.
However, the late dynamics from SP to ST are unaffected by the choice of lattice type.
Therefore, the final spatiotemporal patterns and late coarsening are robust to the lattice choice.

\section{Summary}\label{sec:sum}

We clarified the detailed dynamics of the six-state active Potts models as follows:
(i) The difference between DS and SP waves, comprising all six states, was clarified.
Shallow peaks or minima appear in the contact probability, correlation length, and correlation time at the boundary between these two modes.
(ii) TG and ST wave modes were newly observed under strong repulsion at nonflip contacts.
These modes can coexist via hysteresis.
During coarsening from a random mixture to ST waves, SP waves form temporally as intermediate states.
(iii) The transition from wave to mixing modes is also characterized by the time autocorrelation function.
(iv) Previously, an SP wave comprising three states was considered a single mode.
However, these waves comprise forward and backward waves, which propagate in opposite directions. 
(v) The transition between the waves comprising even- and odd-numbered states is first-order under asymmetric contact energies, similar to the Ising model under a magnetic field.
Moreover, we confirmed that these dynamics occur not only in square lattices but also in the hexagonal lattices.

In this study, we expanded the simulation conditions to include stronger repulsion at nonflip contacts.
However, the parameter space has not yet been fully covered.
When stronger attractions are used at nonflip contacts than at contacts between the same states ($J_{\mathrm{nf}}>J_{k,k}$),
we found a local alternative arrangement in a checkerboard pattern, similar to that in antiferromagnetism.
The local arrangement can modify wave dynamics.\cite{nogu26d}
Several modes of the eight-state cyclic Potts model and in competitive flip cycles have been investigated in Refs.~\onlinecite{nogu25b} and \onlinecite{nogu26a}, 
respectively.
Other dynamic modes can emerge under unexplored conditions, such as larger numbers of states and different flip networks.
Overall, the active Potts models provide an excellent platform for further exploration of spatiotemporal patterns under thermal fluctuations.

\begin{acknowledgments}
The simulations were partially carried out at ISSP Supercomputer Center, University of Tokyo (ISSPkyodo-SC-2025-Ca-0049).
This work was supported by JSPS KAKENHI Grant Number JP24K06973. 
We thank Hajime Tanaka for his lectures and discussions on coarsening dynamics and other topics,
which have stimulated our studies.
\end{acknowledgments}

\section*{DATA AVAILABILITY}
The data that support the findings of this article are openly available.\cite{datarep}

\end{document}